\documentclass[12pt,a4paper]{article}

\newcommand{\be}{\begin{equation}}
\newcommand{\ee}{\end{equation}}
\newcommand{\bea}{\begin{eqnarray}}
\newcommand{\eea}{\end{eqnarray}}

\newcommand{\Tr}{\mathop{\mathrm{Tr}}\nolimits}
\newcommand{\dd}{\mathrm{d}}

\begin{document}

\title{Bilocal Dynamics in Quantum Field Theory}

\author{Ciprian Acatrinei\thanks{On leave from: {\it Institute of
        Atomic Physics  -
        P.O. Box MG-6, 76900 Bucharest, Romania}; e-mail:
        acatrine@physics.uoc.gr.} \\
        Department of Physics, University of Crete, \\
        P.O. Box 2208, Heraklion, Greece}

\date{September 15, 2002}

\maketitle

\begin{abstract}
An essential aspect of noncommutative field theories is their bilocal nature.
This feature, and its role in the IR/UV mixing, are discussed 
using a canonical quantization procedure developed recently.
\end{abstract}

Locality has been long considered as an essential ingredient
of Quantum Field Theories, although attempts to go beyond this
powerful constraint ocasionally appeared.
Recently, a peculiar form of nonlocality atracted attention,
in the context of noncommutative (NC) field theories (FT) \cite{reviews}.
Intuitive, stringy or Weyl-Moyal based arguments appeared to favour
a dipolar nature of the degrees of freedom of such theories \cite{dipoles}.

We will present here a different approach, based on a canonical quantization procedure
developed recently \cite{a1}. 
It clearly demonstrates the intrinsic bilocal  nature of noncommutative fields, 
and renders transparent 
the nature of the real space-time on which dynamics takes place, and on which
measurements could be performed (as opposed to the fictitious Weyl symbols space).
This approach allows one to view our space from
different perspectives \cite{a1,a2}, corresponding to the representation
of the NC algebra one chooses. 
Comments on the IR/UV mixing are also presented.

{\bf Bilocal objects}

The simplest NC field is a $(2+1)$-dimensional scalar  $\Phi(t,\hat{x},\hat{y})$,
defined over a commuting time $t$ and a pair of NC coordinates which satisfy
\be
[\hat{x},\hat{y}]=i\theta. \label{nc}
\ee
The extension to several NC pairs is straightforward. The action is 
\be
S=\frac{1}{2}\int\dd t \Tr_{{\cal H}} 
\left [
\dot{\Phi}^2
-(\partial_x \Phi)^2 -(\partial_y \Phi)^2
-m^2\Phi^2- 2 V(\Phi)
\right ].
\label{action}
\ee 
We will exemplify with a quartic potential, $V(\Phi)=\frac{g}{4!}\Phi^4$.   
The operators $\hat{x}$ and $\hat{y}$ act on a harmonic oscillator Hilbert space ${\cal H}$  
in the usual way. ${\cal H}$ may be given 
a discrete basis $\{|n>\}$ formed by eigenstates of $\hat{x}^2+\hat{y}^2$ \cite{a2},
or a continuous one $\{|x>\}$, composed of eigenstates of, say, $\hat{x}$ \cite{a1}. 

To quantize $\Phi$ \cite{a1}, start with a usual classical commuting field,
expanded into normal modes with coefficients $a$ and $a^{*}$.
Upon usual field quantization, $a$ and $a^{*}$
become operators acting on a standard Fock space ${\cal F}$.
To make the underlying space noncommutative, introduce (\ref{nc})
and apply the Weyl quantization procedure \cite{weyl}
to the exponentials  $e^{i(k_x x+k_y y)}$. The result is 
\be
\Phi=\int\int\frac{dk_x dk_y}{2\pi\sqrt{2\omega_{\vec{k}}}}
\left [
\hat{a}_{k_x k_y}e^{i(\omega_{\vec{k}}t-k_x\hat{x}-k_y\hat{y})}
+\hat{a}^{\dagger}_{k_x k_y}e^{-i(\omega_{\vec{k}}t-k_x\hat{x}-k_y\hat{y})}
\right ].
\label{Qf}
\ee
which means the following: 
$\Phi$ is a `doubly'-quantum field operator, 
acting on a direct product of two Hilbert spaces,
$\Phi :{\cal F}\otimes {\cal H}\rightarrow {\cal F}\otimes {\cal H}$.
Physically, $\Phi$ creates (destroys), 
via $\hat{a}^{\dagger}_{k_xk_y}$ ($\hat{a}_{k_xk_y}$), an excitation 
represented by a "plane wave" 
$e^{i(\omega_{\vec{k}}t-k_x\hat{x}-k_y\hat{y})}$.
The nature of such an excitation will be discussed now.
%We will now describe such an object in familiar space-time terms. 

One could work with $\Phi$ as an operator ready to act on both 
${\cal F}$ and ${\cal H}$ .
It is however simpler to saturate its action on ${\cal H}$, 
working with expectation values
$<x'|\Phi|x> :{\cal F}\rightarrow {\cal F}$. 
It is at this point, of eliminating noncommutativity, that bilocality appears.
To see that, consider the family $\{|x>\}$ of eigenstates
of $\hat{x}$: $\hat{x}|x>=x|x>$,
$\hat{y}|x>=-i\theta\frac{\partial}{\partial x}|x>$.
A simple but key equation is
\be
<x'|e^{i(k_x\hat{x}+k_y\hat{y}}|x>
=e^{ik_x(x+k_y\theta/2)}\delta(x'-x-k_y\theta)
=e^{ik_x\frac{x+x'}{2}}\delta(x'-x-k_y\theta). \label{bilocal}
\ee
This is a bilocal expression, 
and we already see that its span along the $x$ axis,
$(x'-x)$, is proportional to the momentum along  the conjugate $y$ direction, 
i.e. $(x'-x)=\theta k_y$.
In general, for $n$ pairs of NC directions, 
one can  keep only one coordinate out of every pair;
commutativity is gained on the reduced space, at the expenses of strict locality.
Using (\ref{Qf},\ref{bilocal}), one sees that 
\be
<x'|\Phi|x>=\int \frac{dk_x}{2\pi\sqrt{2\omega_{k_x,k_y}}}
\left [
\hat{a}_{k_x, k_y}e^{i(\omega_{\vec{k}}t-k_x\frac{x+x'}{2})}
+\hat{a}^{\dagger}_{k_x ,-k_y}e^{-i(\omega_{\vec{k}}t+k_x\frac{x+x'}{2}}
\right ]
\label{bif}
\ee
where $k_y=(x'-x)/\theta$.
Thus, $\Phi$ annihilates  a rod of (arbitrary) momentum $k_x$ 
and (fixed) length $\theta k_y$,
and creates a rod of momentum $k_x$ and length $-\theta k_y$.
Due to (\ref{nc}), one degree of freedom apparently disappears from (\ref{bif}).
Its presence shows up only through the modified dispersion relation 
\be
\omega_{(k_x, k_y=\frac{x'-x}{\theta})}=\sqrt{k_x^2+\frac{(x'-x)^2}{\theta^2}+m^2}.\label{energy}
\ee
One notices the intrinsic IR/UV-dual character of the dipoles:
both big momentum (UV) and big extension (IR) increase the energy.
This second term reminds a string stretched between two separated D-branes.

Other bases can also be used for ${\cal{H}}$. 
For instance, the basis $\{|n>\}$, formed by the eigenvectors of  $\hat{n}\sim x^2+y^2$,
leads to a discrete remnant space \cite{a2}.

{\bf Correlators}

Two-point correlation functions for such dipoles are 
the VEV of the product of two bilocal fields
(taken on the vacuum, $|0\rangle$, of the Fock space ${\cal F}$):
\be
\langle 0| <x_4|\Phi|x_3> <x_2|\Phi|x_1> |0\rangle =
\int \frac{dk_x}{8\pi^2 \omega_{\vec{k}}}
e^{ik_x[\frac{x_3+x_4}{2}-\frac{x_1+x_2}{2}]}\delta(x_4-x_3-x_2+x_1).
\label{propagator}
\ee
Again, $k_y=(x'-x)/\theta$, 
$\omega_{\vec{k}}=\omega_{k_x,k_y}$ obeys (\ref{energy}),
and there is no integral along $k_y$. 
If one compares (\ref{propagator}) to the $(1+1)$-dimensional 
correlator of two commutative fields,
$\langle0|\phi(X_2)\phi(X_1)|0\rangle$, with $X_1=(x_1+x_2)/2$ and  $X_2=(x_3+x_4)/2$,
the differences are
the $\frac{(x'-x)^2}{\theta^2}$ term in (\ref{energy}), 
and the delta function $\delta([x_4-x_3]-[x_2-x_1])$,
which ensures that the length of the rod is conserved.
Thus, our bilocal objects propagate in a $(1+1)-$dimensional space. 
The extra $y$ direction
is accounted for by their lenght, which contributes to the energy, 
and their orientation. Although we also call these rods dipoles,
they do not necessarily have charges at their ends
and they have extension in the absence of any background.
Those rods may remind one about stretched open strings,
or the double index representation of Yang-Mills theories.

{\bf Interactions}

The quartic interaction term in (\ref{action}) can be written as
\be
\int dt Tr_{{\cal{H}}} V(\Phi)
=\frac{g}{4!}\int dt \int_{x,a,b,c}
<x|\Phi|a><a|\Phi|b><b|\Phi|c><c|\Phi|x>.
\label{potential}
\ee
To find the Feynman rules, we need the vacuum correlator (\ref{propagator}),
and a slight modification of the Dyson procedure.
The basic `vertex' for four-dipole scattering follows from
\be
\langle -\vec{k}_3, -\vec{k}_4|
:\int dt \int_{x,a,b,c}
<x|\Phi|a><a|\Phi|b><b|\Phi|c><c|\Phi|x> :
|\vec{k}_1, \vec{k}_2 \rangle .
\label{vertex}
\ee
$|\vec{k}_1, \vec{k}_2 \rangle$ is a Fock space state with two quanta 
of momentum $\vec{k}_1$ and $\vec{k}_2$.
The momenta $\vec{k}_{i, i=1,2,3,4}$ have each two components: $\vec{k}_i=(k_i,l_i)$.
$k_i$ is the momentum along $x$, whereas $l_i$ represents the dipole
extension along $x$ (corresponding to the momentum along $y$) . 
Using Eq. (\ref{bif}) and integrating over $x,a,b$ and $c$,
one obtains the conservation laws $k_1+k_2+k_3+k_4=0$ and $l_1+l_2+l_3+l_4=0$.
%With their help  one sees that, 
The final result differs from the  four-point scattering
vertex of $(2+1)$ commutative particles with  momenta $\vec{k}_i=(k_i,l_i)$
only through the phase
\be
e^{-\frac{i\theta}{2}\sum_{i<j}(k_i l_j-l_i k_j)}. \label{phase}
\ee
%Interpreting $l_i$ as the $i$-th momentum along $y$, 
This is precisely the star-product modification of the usual Feynman rules.
The phase (\ref{phase}) appears due to the bilocal nature
of generic $<x'|\Phi|x>$'s. 
%Pointlike $<x|\Phi|x>$'s would never produce it.

By contracting various terms in (\ref{vertex}), one obtains the 
one-loop corrections to the free rod propagator,
together with the recipe for calculating loops. 
Again, the derivation is straightforward.
%, hence we do not reproduce it.
The main point is that, in the end,
one has to integrate over both the momentum and length of the dipole circulating
in a loop. This $\frac{1}{2\pi}\int dk_{loop} \int dl_{loop}$ integration,
together with the dispersion relation (\ref{energy}),
brings back into play - especially as far as divergences are concerned - the $y$
direction.
It is easy to extend the above reasoning to $(2n+1)-$dimensions:
unconstrained  dipoles will propagate in a $(n+1)$-dimensional commutative space-time,
with Feynman rules obtained as outlined above. 
Once the dipole lengths are interpreted as momenta in the conjugate directions, 
the rules are identical to those obtained long ago  via star-product calculus.

{\bf IR/UV}

We have derived directly from field theory the dipolar character of NC excitations;
the momentum in the conjugate direction became the lenght of the dipole.
A connection between UV and IR physics appeared naturally, 
and on a somehow more rigorous basis than in \cite{ir_uv_2}, for instance. 

One can also view geometrically
the differences between planar and nonplanar loop diagrams,
and the role of low momenta in nonplanar graphs.
To illustrate this, consider $(4+1)$-dimensions, $t,\hat{x},\hat{y},\hat{z},\hat{u}$, 
with $[\hat{x},\hat{y}]=[\hat{z},\hat{w}]=i\theta$. In the $\{|x,z>\}$ basis,
one has a commutative space spanned by the axes $x$ and $z$, 
on which dipoles with momentum $\vec{p}=(p_x,p_z)$ and length
$\vec{l}=(l_x,l_z)=\theta(p_y,p_w)$ evolve. During the scattering, four such dipoles
meet in a four-edged poligon of  area ${\cal A}$
(figure 1a).

\begin{picture}(350,230)(10,-20)
\thicklines
%figure 2a
\put(40,110){\vector(1,1){40}}
\put(80,150){\vector(1,-1){50}}
\put(130,100){\vector(-2,-1){74}}
\put(56,63){\vector(-1,3){16}}
\put(32,173){\vector(2,-3){15}}
\put(120,135){\vector(4,3){20}}
\put(100,76){\vector(4,-1){20}}
\put(30,35){\vector(1,3){12}}
%figure 2b
\put(230,140){\vector(-1,2){20}}
\put(212,182){\vector(1,-2){20}}
\put(250,130){\vector(-2,1){20}}
\put(232,142){\vector(2,-1){20}}
%figure 2c
\put(220,85){\vector(2,1){30}}
\put(250,100){\vector(1,-4){10}}
\put(260,60){\vector(-2,-1){30}}
\put(230,45){\vector(-1,4){10}}
%figure 2d
\put(320,90){\vector(2,1){4}}
\put(324,92){\vector(1,-4){10}}
\put(334,52){\vector(-2,-1){4}}
\put(330,50){\vector(-1,4){10}}
%\put(67,105){\bf ${ \cal A}\neq 0$}
\thinlines
\put(40,110){\line(-2,3){20}}
\put(80,150){\line(-2,3){20}}
\put(80,150){\line(4,3){30}}
\put(130,100){\line(4,3){30}}
\put(130,100){\line(4,-1){30}}
\put(56,63){\line(4,-1){30}}
\put(56,63){\line(-1,-3){15}}
\put(40,110){\line(-1,-3){15}}
\put(67,105){${\cal A}\neq 0$}
\put(225,65){${\cal A}\neq 0$}
\put(250,160){${\cal A}= 0$}
\put(340,70){${\cal A}= 0$}
\put(150,140){\vector(1,0){50}}
\put(155,150){planar}
\put(152,55){nonplanar}
\put(150,70){\vector(1,0){50}}
\put(270,70){\vector(1,0){40}}
\put(270,55){$\vec{l}_{ext}\rightarrow 0$}
\put(221,165){$\vec{l}_{ext}$}
\put(220,100){$\vec{l}_{ext}$}
\put(240,140){$\vec{l}_{loop}$}
\put(256,88){$\vec{l}_{loop}$}
\put(42,122){$\vec{l}_{1}$}
\put(49,89){$\vec{l}_{2}$}
\put(70,78){$\vec{l}_{3}$}
\put(94,139){$\vec{l}_{4}$}
\put(39,169){$\vec{k}_{1}$}
\put(20,40){$\vec{k}_{2}$}
\put(94,59){$-\vec{k}_{3}$}
\put(119,152){$-\vec{k}_{4}$}
\put(80,30){(a)}
\put(240,30){(c)}
\put(225,186){(b)}
\put(330,30){(d)}
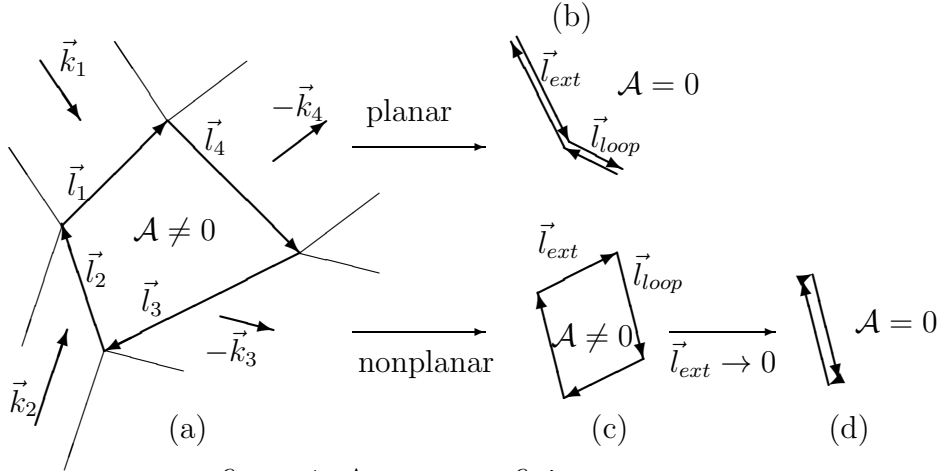
\put(100,10){figure 1: Area versus finiteness}
\end{picture}

\noindent
One has two possibilities for the one-loop correction to the propagator:
planar and nonplanar.
In the planar case, adjacent dipole fields are contracted. Momentum and
length conservation enforce then the poligon to degenerate into a
one-dimensional, zero-area object (figure 1b). UV divergences persist.
In the nonplanar case, due to the nonadjacent contraction
the area ${\cal A}$ does not go to zero (cf. figure 1c)
unless the external dipole length vanishes (figure 1d). 
${\cal A}\neq 0$ appears thus to be related to the disappearance of UV divergences.
Actually, the true regulator is the phase (\ref{phase}).
This is zero, i.e. ineffective, when ${\cal A}= 0$
in {\it both} the $|x,z>$ and $|y,u>$ bases.
That corresponds to zero external length {\it and} momentum
in the dipole picture, which means that the resulting divergence is 
half IR ($\vec{p}_{ext}=0$) and half UV ($\vec{l}_{ext}=0$)!
In Weyl space this is just the usual zero external momentum,
say $p_{\mu}^{ext}=0$, and one speaks about an IR divergence.
For dipoles the divergence comes from having zero vertex area ${\cal A}$ in any basis,
and is half IR and half UV.
NCFT appears to be somehow in between usual FT and string theory:
when the interaction vertex is a point, UV infinities appear;
when it opens up, as in string theory, amplitudes are finite. 

{\bf Remarks}

We saw that by dropping $n$ coordinates, intuition is gained:
the remaining space admits a notion of distance,
although bilocal (and in some sense IR/UV dual) objects probe it.
%Other bases of ${\cal{H}}$ can also be used. 
%For instance, the basis $\{|n>\}$, formed by eigenvectors of  $\hat{n}\sim x^2+y^2$,
%leads to a discrete remnant space \cite{a2}.
An important  question is: how do the dimensionality and noncommutativity of space-time
exactly depend on the regime in which we probe the theory?
To start, we have a NC $(2n+1)-$dimensional theory. Then, at tree level
(i.e. classical plus tree level interference effects), 
one has $D=n+1$ commuting directions. 
However, loop effects drive us back to $D=2n+1$.
At a scale $r\sim \sqrt{\theta}$, space is NC. 
For  $r>> \sqrt{\theta}$ it is believed to be commutative. 
However,
if $r$ is the radius in the largest available commutative subspace,
the IR/UV connection 
suggests a connection (duality?) between the $r>> \sqrt{\theta}$ and $r<< \sqrt{\theta}$ regimes.
A clarification of these issues is desirable.

One may also consider the case in which time is NC, e.g.
$[\hat{t},\hat{x}]\neq 0$. In a basis in which $\hat{t}$ is diagonal,
$\{|t,\dots>\}$, the elementary excitations become bilocal in time, 
$<t,\dots|\Phi|t',\dots>$.
Their time-length contributes to the energy, 
$\omega=\sqrt{(t-t')^2/\theta^2+k_x^2+k_y^2+m^2}$. 
Preliminay results indicate that,
upon appropriate definition of the perturbation series, the theory {\it is} unitary,
in agreement with \cite{t_unitary} and in disagreement with \cite{t_nonunitary}.

\vskip 0.3cm

{\bf Acknowledgments} 

This work was supported through a European Community Marie Curie fellowship,
under Contract HPMF-CT-2000-1060.

%%%%%%%%%%%%%%%%%%%%%%%%%%

\end{document}